\documentclass[aps,prf,groupedaddress,twocolumn,floatfix]{revtex4-2}
\usepackage{textcomp}
\usepackage{graphicx}
\usepackage{subfigure}
\usepackage{natbib}
\usepackage{color}
\usepackage{bm}
\usepackage{amssymb}
\usepackage{amsmath}
\usepackage{amsthm}
\usepackage[dvipsnames]{xcolor}

\begin{document}

\title{Extreme vorticity events in turbulent Rayleigh-B\'enard convection from stereoscopic measurements and reservoir computing}
\author{Valentina Valori}
\affiliation{Institute of Thermodynamics and Fluid Mechanics, Technische Universit\"at Ilmenau, D-98684 Ilmenau, Germany}
\author{Robert Kr\"auter}
\affiliation{Institute of Thermodynamics and Fluid Mechanics, Technische Universit\"at Ilmenau, D-98684 Ilmenau, Germany}
\author{J\"org Schumacher}
\affiliation{Institute of Thermodynamics and Fluid Mechanics, Technische Universit\"at Ilmenau, D-98684 Ilmenau, Germany}
\affiliation{Tandon School of Engineering, New York University, New York, NY 11201, USA}

\date{\today}

\begin{abstract}
High-amplitude events of the out-of-plane vorticity component $\omega_z$ are analyzed by stereoscopic particle image velocimetry (PIV)  in the bulk 
region of turbulent Rayleigh-B\'{e}nard convection in air. The Rayleigh numbers ${\rm Ra}$ vary from $1.7 \times 10^4$ to $5.1 
\times 10^5$. The experimental investigation is connected with a comprehensive statistical analysis of long-term time series of $\omega_z$ and individual 
velocity derivatives $\partial u_i/\partial x_j$. A statistical convergence for derivative moments up to an order of 6 is demonstrated. Our results are found to 
agree well with existing high-resolution direct numerical simulation data in the same range of parameters, including the extreme vorticity events which 
appear in the far exponential tails of the corresponding probability density functions. The transition from a Gaussian to a non-Gaussian velocity derivative 
statistics in the bulk of a convection flow is confirmed experimentally. {The experimental data are used to train a reservoir computing model, one 
implementation of a recurrent neural network, to reproduce highly intermittent experimental time series of the vorticity and thus reconstruct extreme 
out-of-plane vorticity events. After training the model with high-resolution PIV data, the machine learning model is run with sparsely seeded, continually 
available, and unseen measurement data in the reconstruction phase. The dependence of the reconstruction quality on the sparsity of the partial 
observations is also documented.} Our latter result paves the way to machine--learning--assisted experimental analyses of small-scale turbulence for 
which time series of missing velocity derivatives can be provided by generative algorithms.
\end{abstract}

\maketitle

\section{Introduction}
Thermal convection is one the fundamental mechanisms by which the turbulent flows evolve. The temperature dependence of the fluid mass density and the resulting buoyancy forces drives fluid motion that in turn advects the temperature leading to a fully turbulent motion of the fluid \cite{Kadanoff2001,Ahlers2009,Chilla2012,Verma2018}. The simplest configuration in this specific class of turbulent flows is Rayleigh-B\'{e}nard convection (RBC) -- a fluid layer of height $H$ between two parallel rigid plates heated uniformly from below and cooled uniformly from above, such that a constant temperature difference $\Delta T=T_{\rm bot}-T_{\rm top}>0$ is sustained across the layer. Experimentally, RBC can be realized in a horizontally extended closed cylindrical or cuboid cell with thermally insulated sidewalls. In the past two decades, a larger number of direct numerical simulations (DNS) of this configuration investigated various aspects of the large-scale structure formation, the longer-term dynamics and the dependencies on Rayleigh and Prandtl numbers, ${\rm Ra}$ and ${\rm Pr}$, as well as on the aspect ratio $\Gamma$ of the cells in detail \cite{Hartlep2003,Parodi2004,Hartlep2005,Hardenberg2008,Bailon2010,Emran2015,Pandey2018,Stevens2018,Fonda2019,ValoriDubrulle2019,Green2020,Krug2020,Berghout2021}. The number of studies of RBC in air at a Prandtl number of ${\rm Pr}=0.7$ by means of controlled laboratory experiments is much smaller, see e.g. refs. \cite{Deardorff1967,Fitzjarrald1976,Schmeling2014,Kaestner2018,Cierpka2019} for investigations in large-aspect-ratio setups. The three dimensionless control parameters of RBC experiments are defined as follows,
\begin{equation}
{\rm Ra}=\frac{g\alpha \Delta T H^3}{\nu\kappa}\,,\quad {\rm Pr}=\frac{\nu}{\kappa}\,,\quad \Gamma=\frac{L}{H}\,.
\end{equation}
Here, $g$ is the acceleration due to gravity, $\alpha$ the thermal expansion coefficient, $\nu$ the kinematic viscosity, $\kappa$ the thermal diffusivity, and $L$ the horizontal length scale.

The dynamics and structure of the boundary layers of the velocity and temperature fields at the top and bottom of an RBC layer are essential for the amount of heat that can be carried from the bottom to the top in this configuration. In Valori and Schumacher \cite{Valori2021}, a dynamical connection between this dynamics and the small-scale intermittent motion in the bulk, in particular the formation of high-amplitude dissipation events, was established. Thermal and kinetic energy dissipation fields probe the magnitude of gradients of the temperature and velocity fields, respectively. Their amplitudes are known to be largest at the smallest scales \cite{Schumacher2010,Scheel2013,Yeung2015,Buaria2020}. In detail, it was shown in \cite{Valori2021} how the formation of coherent plumes at the top and bottom and their subsequent collision or passing creates large temperature gradients in the bulk that cause the formation of localized shear layers, see also ref. \cite{Scheel2016} for a closed cylindrical cell at unit aspect ratio. These studies set one motivation for the present work.

In the present work, we want to study the velocity derivatives in the bulk of an RBC configuration experimentally by means of stereoscopic particle image velocimetry (SPIV) in the midplane of the horizontal cell. This allows us to obtain all three velocity components $u_i({\bm x},t)$ in a horizontal measurement region $A=2.9 H\times 2.2 H$ and thus  7 out of the 9 components of the velocity gradient tensor field $M_{ij}({\bm x},t)=\partial u_i/\partial x_j$. These are the in-plane component $\partial u_i/\partial x_j$ with $i=1,2,3$ and $j=1,2$ and $\partial u_3/\partial x_3=-(\partial u_1/\partial x_1+\partial u_2/\partial x_2)$ via the incompressibility condition of the flow. Thus the out-of-plane component of the vorticity field $\omega_3=\partial u_2/\partial x_1-\partial u_1/\partial x_2$ can be obtained and we have access to highly intermittent derivative fields in the bulk of the turbulent convection layer. We analyze the statistics of these 7 derivatives and probe their statistical convergence. Furthermore, the resulting probability density functions (PDFs) are found to agree to those of existing high-resolution DNS data from ref. \cite{Valori2021} in a similar setup and the same parameter range. 

The SPIV snapshot series contain a few extreme events in the form of intense vortex cores that sweep across the measurement plane $A$. In the second part of this work, we analyze their temporal growth and explore the capability of recurrent neural network (RNN) architectures to 
{reconstruct high-amplitude or extreme events of the out-of-plane vorticity component \cite{Goodfellow2016}.} More specifically, we therefore apply the 
reservoir computing model (RCM) \cite{Jaeger2004,Pandey2020} which has been shown to successfully predict the time evolution of nonlinear dynamical 
and fluid mechanical systems. {Examples for the latter case are the Kuramoto-Sivashinsky equation \cite{Lu2017,Vlachas2020a,Vlachas2020b}, 
low-dimensional Galerkin models of wall-bounded shear flows \cite{Srinivasan2019,Eivazi2021}, the two-dimensional Kolmogorov flow 
\cite{Farazmand2021}, or two-dimensional RBC configurations without and with phase changes \cite{Pandey2020a,Heyder2021}. It should be also 
mentioned that other methods for the temporal predictions of complex dynamics are available which do not rely on neural network approaches, such as 
the Koopman framework with nonlinear forcing \cite{Eivazi2021}.} 

{We show that an RCM with continually available sparse data is able to reconstruct high-amplitude or extreme out-of-plane vorticity events, which can 
be quantified by a squared vorticity integrated over $A$ (see ref. \cite{Sapsis2021} for a recent review). In other words, the present machine learning 
algorithm thus operates in an open-loop scenario after the training phase and continues to use previously unseen sparse data. The prescribed data 
amount to a percentage of the full PIV resolution which varies between 1.9\% and 11\%.} 

The outline of the manuscript is as follows. In Sections IIA to IIC, details on the experiment and the machine learning algorithm are provided. Section IIIA 
discusses the results in respect to the vorticity and velocity derivative statistics. Furthermore, we investigate the dynamics of a particular high-vorticity 
event tracked in the experiment in Sec. IIIB. The subsequent section IIIC finally provides the results of the application of the RNN to reconstruct  the time 
evolution and the non-Gaussian PDFs with the extended tails. We conclude the work with a summary and an outlook in Section IV. For convenience, we 
will also switch from the notation $u_1$, $u_2$, $u_3$, and $\omega_3$ for velocity and vorticity field components to $u_x$, $u_y$, $u_z$, and 
$\omega_z$ in the following sections with coordinate $z$ parallel to the inferred temperature gradient between top and bottom. 

\begin{figure}
\includegraphics[width=0.4\textwidth]{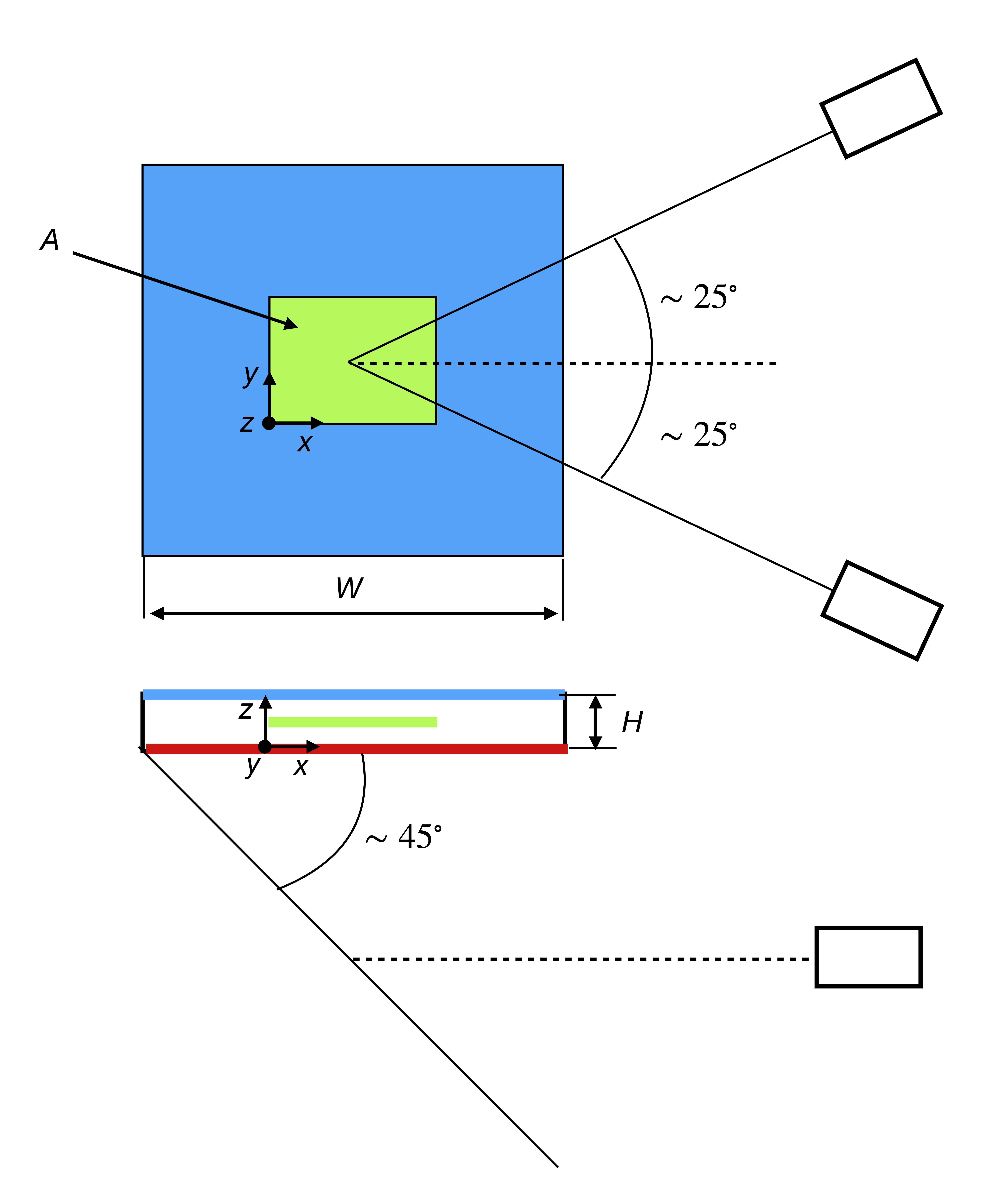}
\caption{Relative position of cameras (shown as rectangles), RB cell and measurement section from the top and side view. The heated bottom plate of the 
RB cell and the cooled top plate are indicated by a red and a blue line, respectively. The measurement section $A$ is represented in green. $W$ and $H$ 
are the horizontal and the vertical dimensions of the cell respectively.\label{fig:RBC-setup}}
\end{figure}

\section{Methods}
\subsection{Rayleigh-B\'{e}nard convection cell for pressurized air}
The experimental set-up is a large--aspect ratio Rayleigh-B\'{e}nard convection (RBC) cell of size $W\times W\times H$, where the horizontal dimension $W$ is ten times larger than the vertical distance between the two plates $H$, the latter being 3 cm (see figure \ref{fig:RBC-setup}). The aspect ratio of the cell is therefore $\Gamma=10$. The bottom plate of the cell is made of two glass plates coated on one side with a thin layer of Indium Tin Oxide. It has the important characteristic of being transparent and at the same time uniformly heatable by the Joule effect. The latter arises from the electrical current that goes through the oxide layer (for further details see  \cite{Kaestner2018}). Each plate has a measured light transmission coefficient of about 68$\%$ \cite{Kaestner2018}. This coating was manufactured at the Fraunhofer Institute for Organic Electronics in Dresden (Germany). The top plate of the RBC cell is made of Aluminium with an internal cooling circuit where water flows at controlled temperature fed by a thermostat. The side walls are made of 4 mm thick polycarbonate that allows optical access for the laser light. The surface temperatures at both horizontal walls were measured by four thermoresistances PT100 (class B) for each plate, which were located 20 mm far from the side walls. 

The RBC cell is inserted in the high-pressure facility, the Scaled Convective Airflow Laboratory Experiment (SCALEX). This facility consists of a high-pressure vessel with 35 observation windows, allowing for optical access from the outside. The pressure within the vessel can be regulated from 10-100 mbars to 8 bars in steps of about 100 mbars. The working pressure was measured with a Cerabar PMC131 sensor from Endress + Hauser AG. Further details on the device are for example summarized in ref. \cite{Kaestner2018}.

\subsection{Particle Image Velocimetry measurements}
Stereoscopic particle image velocimetry experiments are performed in a horizontal layer at mid height between top and bottom plate of the cell. The position of the measurement plane is shown in green in figure \ref{fig:RBC-setup}. The measurement region covers a horizontal area of $A\approx 2.9 H\times 2.2 H$. In order to conduct SPIV measurements in a horizontal plane within the cell, two cameras are located below the cell which take images through the transparent bottom plate. A mirror is placed in the optical path as represented in figure \ref{fig:RBC-setup}. A stereo angle of about $50^\circ$ is used. The cameras are sCMOS from LaVision GmbH with digital resolution of $2560 \times 2160$ pixels and a pixel pitch of $6.5 \mu$m. Therefore the image magnification factor is $M \sim0.2$. The cameras have particularly thin connector cables made of optical fibres that allow them to be placed inside the high-pressure vessel. These cables are inserted into feed through systems for high pressure from Spectite$\textsuperscript{\tiny{\textregistered}}$ that are built in a flange of the vessel. The images are recorded at a frame rate of 10 Hz with interframe time of 7 ms or 10 ms depending on the strength of the convective flow (or Rayleigh number ${\rm Ra}$) studied. Each camera is equipped with a Zeiss Milvus 2/100M objective lens under the Scheimpflug condition. The focal length of the lenses is 100 mm and the aperture stop used is 5.6 and 8 for the camera in forward and backward scattering, respectively. Both cameras and lenses are placed inside the vessel at a working pressure of up to 4.5 bars. The laser light sheet was created with a double pulse laser (Quantel Q-smart Twins 850) with a pulse energy of about 175 mJ for a 2 mm thick light sheet. A spherical and a cylindrical lens were used to generate the light sheet.

 \begin{table}
 \begin{ruledtabular}
 \begin{tabular}{ccccccc}
 	Ra &  $T_{\rm top} [^{\circ}$C$]$ & $T_{\rm bot} [^{\circ}$C$]$ &$\Delta T$ & p [bars] & $U_{\rm ff}$ [mm/s] & $T_{\rm ff}$ [s]\\
 \hline
 	$1.7\times10^4$  & $22.3$ & $28.9$ & $6.6$  &1            & $80.7$ & $0.4$\\
 	$2.1\times10^4$  & $21.2$ & $29.2$ & $8$     &1           & $80.7$ & $0.4$\\
	$1.1\times10^5$  & $21.2$ & $28.5$ & $7.3$  & $2.47$  & $85$    & $0.35$\\
    $2.9\times10^5$  & $22.9$ & $28.4$ & $5.5 $ & $4.5$    & $74.3$ & $0.4$\\
    $5.1\times10^5$  & $20.1$ & $29.9$ & $9.8$  & $4.5$    & $98.8$ & $0.3$\\
\end{tabular}
\end{ruledtabular}
\caption{Summary of the experimental conditions used in the convection measurements. The first column reports the Rayleigh numbers, the second and 
the third one the values of the top ($T_{\rm top}$) and bottom ($T_{\rm bot}$) temperature, respectively, the fourth one the temperature difference 
between the bottom and the top walls ($\Delta T$), the fifth one the working pressure. The sixth and the seventh columns are for the characteristic 
free-fall velocity and free-fall time of the flow, respectively.\label{tab:expProgramme}}
\end{table}

The depth of field of the measurements $\delta_z$ is given by
\begin{equation}
\delta_z =4(1+\frac{1}{M})^2f_\#^2\lambda\,
\label{eq:depth-of-field} 
\end{equation}
where $\lambda$ is the wavelength of the laser. Here, $\lambda = 532$ nm for the green laser is used. For the camera in forward scattering with the 
smallest f-stop $f_\#$, the depth is $\delta_z = 2.4$ mm, which is larger than the laser sheet thickness and therefore ensures good focusing conditions of 
the illuminated particles. The geometrical calibration is made on the measurement plane with the three-dimensional target 204-15 from LaVision, where 
the distance between two dots is 15 mm, the dot diameter and the level separation are both 3 mm. Additionally, a stereoscopic self-calibration 
\cite{Wieneke2005} is made and iteratively repeated with final mean residual displacement below 1 pixel. {This final value cannot be further reduced using 
images of a thermal convection flow because of the presence of optical distortions due to variations in the mass density of the imaging medium 
\cite{Valori2018,Valori2019,Valori2021b}.}

\begin{table}
\begin{ruledtabular}
\begin{tabular}{cccc}
			${\rm Ra}^{\rm DNS}$ & $\eta_K$ [mm]  & ${\rm Ra}^{\rm PIV}$ &  $\Delta x^{\rm PIV}$ [mm]\\
			\hline
			$1.5\times10^4$ & $2.4$ & $1.7\times10^4$ & 4\\
			$2\times10^4$ & $ 2.2$  &  $2.1\times10^4$ & 4\\
			$1\times10^5$ & $1.3$  & $1.1\times10^5$ &4\\
			$2\times10^5$ & $1.0 $ & $2.9\times10^5$ &4\\
			$5\times10^5$ & $0.7$ & $5.1\times10^5$ &3.5\\		
\end{tabular}
\end{ruledtabular}
\caption{Kolmogorov length scale from the direct numerical simulations (DNS) and spatial resolution of the PIV measurements computed without overlap for each of the five Rayleigh numbers studied. The simulations were conducted in a domain of $\Gamma=8$ with periodic boundary conditions at the sidewalls at Pr $=1$ by a spectral element method, see ref. \cite{Valori2021} for details. \label{tab:DNS}}
\end{table}

The seeding in the PIV measurements is established by droplets of di(2ethylhexyl)sebacate (DEHS) with average diameter of 0.9 $\mu$m generated by a vaporizer from PIVTECH GmbH that is placed inside the pressure vessel. The characteristic velocity of the flow is estimated as the free fall velocity $U_{\rm ff} = \sqrt{\alpha g \Delta T H}$, see also table \ref{tab:expProgramme} for the values of $U_{\rm ff}$ and $\Delta T$. The sedimentation velocity of the particles of diameter $d_p$, which are used here as tracers, is given by
\begin{equation}\label{eq:sedimentation}
u_S = \frac{d_p^2(\rho_p-\rho_f)}{18\mu g}\,,
\end{equation}
where $\rho_f$ and $\mu=\rho_f\nu$ are, respectively, the density and the dynamic viscosity of the fluid at the mean temperature and pressure of the 
cell, and $\rho_p$ is the mass density of the particles. The sedimentation velocity of the particles used is negligible as visible by the ratio $u_{S}/U_{\rm 
ff}$ that is about $4\cdot 10^{-6}$ or smaller for all flow conditions studied here. {The particles faithfully follow the flow as indicated by the values of the 
dimensionless Stokes number, ${\rm St}$, defined as the ratio between the characteristic time of a suspended particle and the characteristic time of the 
flow, 
\begin{equation}\label{eq:Stokes}
{\rm St} = \frac{d_{p}\rho_{p}U_{\rm ff}}{18\mu g}\,.
\end{equation}
Values of the Stokes number in the present study are always ${\rm St}< 10^{-2}$, as requested for good tracers particles \cite{PIVbookJerry,PIVbookDLR}.}

The PIV images are acquired and processed with the software DaVis 10 from LaVision. The only preprocessing that is applied to the raw images is a time-averaged image subtraction in order to reduce the noise. A two-passes cross-correlation algorithm with decreasing interrogation window size is used. The size of the final pass is 128 $\times$ 128 pixels for all measurements except for the largest Rayleigh number where the size is 96 $\times$ 96,  which leads to a spatial resolution of about 4 mm and 3.5 mm, respectively (see table \ref{tab:DNS}). Here, we also list the Kolmogorov length $\eta_K$ of the DNS (at the comparable Rayleigh numbers) which are  given in dimensionless units by
\begin{equation}
\eta_K=\left(\frac{\rm Pr}{\rm Ra}\right)^{3/8} \langle\epsilon\rangle_{V,t}^{-1/4}\,,
\end{equation}
where $\langle\epsilon\rangle_{V,t}$ is the combined volume-time average of the kinetic energy dissipation rate field \cite{Scheel2013}. This value is multiplied with the actual cell height $H$ to get a length in physical units as shown in the table.

The values of Rayleigh numbers of the experiments range from ${\rm Ra} = 1.7\times10^4$ to ${\rm Ra} = 5.1\times10^5$ and are listed in table \ref{tab:expProgramme}. The working fluid is air with ${\rm Pr} = 0.7$ for all cases. Rayleigh numbers ${\rm Ra} \ge  1.1\times10^5$ are obtained by putting the vessel under pressure.  In table \ref{tab:expProgramme}, the corresponding values of the working pressure of each experiment are shown together with the values of the temperatures at the bottom ($T_{\rm bot}$) and top ($T_{\rm top}$) walls of the cell in addition to the resulting difference $\Delta T$. The last two columns of the table indicate the characteristic free-fall velocity $U_{\rm ff}$ and resulting free-fall time $T_{\rm ff} = H/U_{\rm ff}$ of the flow. Values of the fluid properties are taken from the NIST RefProp database version 9.1 \cite{NIST}.\\
{The major source of uncertainties in this experimental study is due to density variations within the fluid which are caused by local temperature gradients. 
The relative density differences between the bottom and the top of the cell were up to $3.4$\% for the largest Rayleigh number studied. They caused 
errors in the determination of the three-dimensional particle position and velocity, mostly due to instantaneous and local changes in the magnification 
factor which is connected to the motion of thermal plumes \cite{Valori2018,Valori2019,Elsinga2005}. A global {\em a posteriori} estimation of the relative 
velocity uncertainty of the SPIV measurements based on the correlation statistics method \cite{Wieneke2015} gives an upper bound of $0.6\%$ for the 
experiment at the smallest Rayleigh number and of $2.5\%$ for the one at the largest one.}
\begin{figure*}
\includegraphics[width=0.8\textwidth]{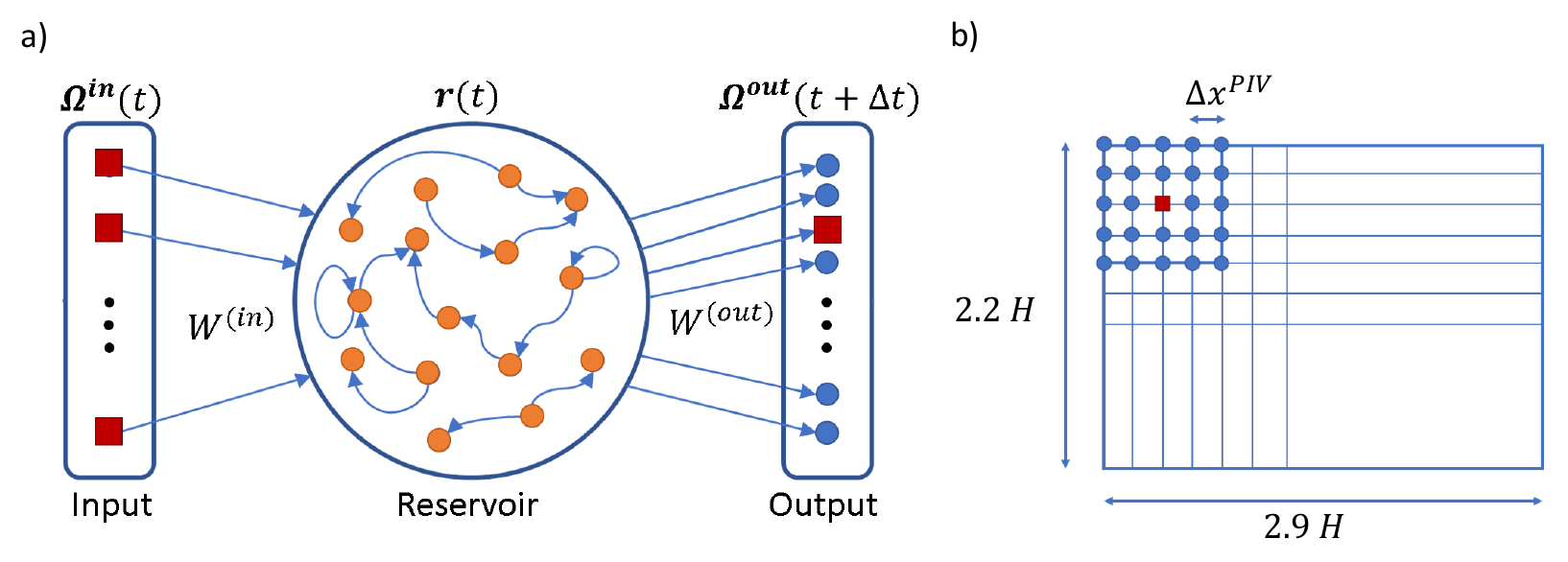}
\caption{Sketch of the reservoir computing model (RCM) and arrangement of continually available data during the prediction phase of the machine 
learning algorithm, indicated as red filled squares. (a) The three building blocks of the RCM are the input layer, the reservoir (a random network of 
neurons), and the output layer. b) Sketch of a part of the data grid $A$ obtained in the stereoscopic particle image velocimetry (PIV) measurement. This 
figure illustrates the $5\times 5$ scenario: one data point with continually available experimental vorticity data (red filled square) is surrounded by 24 grid 
points (blue filled circles) for which the trained RCM predicts the time evolution of the vorticity component autonomously. {Thus 3.5\% of the full PIV 
resolution is available in the reconstruction phase.} The whole measurement area $A$ is covered sparsely with continually available measurements in this 
way. The uniform PIV resolution is also indicated. \label{fig:sketch}}
\end{figure*}
\begin{figure*}
\includegraphics[width=0.85\textwidth]{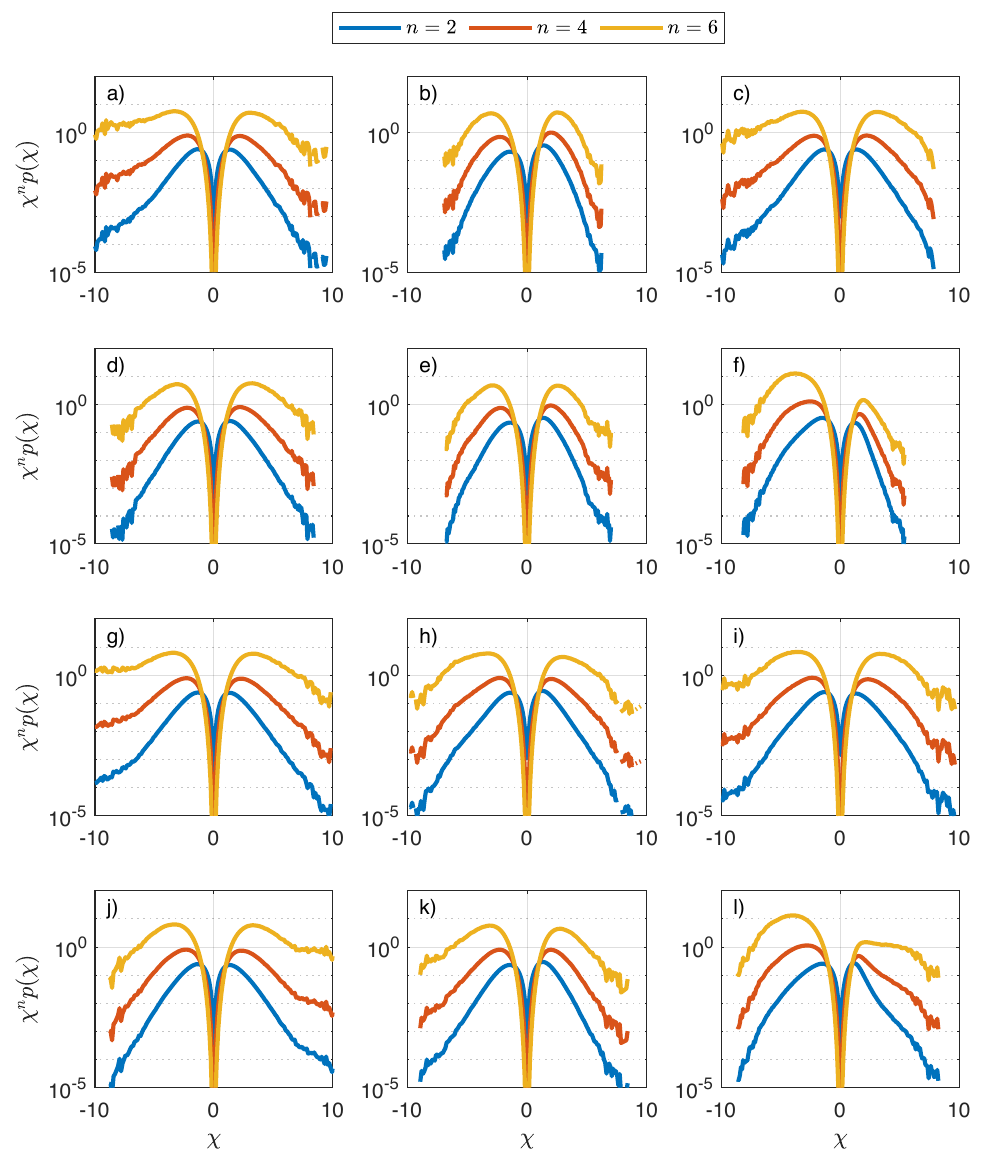}
\caption{Statistical convergence of the higher-order velocity derivative statistics in the experiments. The vorticity component $\omega_z$ (a,g) and the derivatives $\partial u_x/\partial x$ (b,h), $\partial u_y/\partial x$ (c,i), $\partial u_x/\partial y$ (d,j), $\partial u_y/\partial y$ (e,k), as well as $\partial u_z/\partial z$ (f,l) are shown. All derivatives are normalized by their corresponding root mean square values. The moment order $n$ is indicated in the legend above the panels. Panels (a--f) are for ${\rm Ra}=2.9\times 10^5$, panels (g--l) for ${\rm Ra}=5.1\times 10^5$.}
\label{convergence}
\end{figure*}

\subsection{Reservoir computing model}
In the following, we briefly review the reservoir computing model (RCM) \cite{Jaeger2004}. {Our algorithm is a supervised machine learning algorithm 
which is run as a single validation split. This includes (1) training, (2) validation, and (3) test phases. The PIV data are correspondingly split into three 
subsets. Other more complex scenarios, such as $k$-fold cross validations, are possible and have been discussed only recently for RCMs in 
ref.~\cite{Lukosevicius2021}.} 

The reservoir is a random, recurrently and sparsely connected network. The model consists of the input layer, the reservoir, and the output layer as shown in figure \ref{fig:sketch}(a). The input layer takes the training data in the form of discrete time series ${\bm \Omega}^{\rm in}(t)=(\omega_z^1(t),\dots ,\omega_z^{N_{\rm in}}(t))$ with $N_{\rm in}$ the number of selected grid points of the PIV measurement region $A$; see the red squares in figure \ref{fig:sketch}(b). Vorticity data are converted at each instant into a reservoir state vector ${\bm r}(t)\in \mathbb{R}^N$ with a number of reservoir nodes $N\gg N_{\rm in}$. This is done by a random weight matrix  $W^{\rm in} \in \mathbb{R}^{N\times N_{\rm in}}$ which is determined at the beginning of the training and left unchanged, 
\begin{equation}
{\bm r}(t) = W^{\rm in} {\bm \Omega}^{\rm in}(t)\,.
\label{RC0}
\end{equation}
The reservoir is described by a symmetric adjacency matrix $W^r \in \mathbb{R}^{N\times N}$, also determined initially as a random matrix and left 
unchanged. Typically, an ensemble of different random realizations of the reservoir matrix is considered in the training process. Two important parameters 
of $W^r$ are the reservoir density $D$ of active nodes and the spectral radius $\rho(W^r)$, which is set by the largest absolute value of the eigenvalues. 
{The reservoir nodes are updated by a simple nonlinear dynamical system. The discrete time evolution is given by
\begin{align}
{\bm r}(t+\Delta t) &= (1-\alpha){\bm r}(t) \nonumber\\
                            &+\alpha \tanh\left[ W^r {\bm r}(t) + W^{\rm in} {\bm \Omega}^{\rm in}(t)\right]\,,
\label{RC1}
\end{align}
where nonlinearity enters in the form of an activation function, here by a hyperbolic tangent.} The leakage rate is $0<\alpha<1$ and the spectral radius of 
the reservoir is typically taken $\rho(W^r)\lesssim 1$. The final of the three building blocks of the RCM is the output weight matrix $W^{\rm out} \in 
\mathbb{R}^{N_{\rm PIV}\times N}$ which maps the updated reservoir vector back to the vorticity field {and is not random},
\begin{equation}
\hat{\bm \Omega}^{\rm out}(t+\Delta t) = W^{\rm out} {\bm r}(t+\Delta t)\,.
\label{RC2}
\end{equation}
{Here, $N_{\rm PIV}$ is the number of all grid points of the PIV measurement region $A$. The iteration of \eqref{RC1} is repeated for all the snapshots of 
the PIV training data and the sequence of corresponding reservoir states is saved.} In contrast to most other neural networks, the training of the RCM is 
performed with respect to the output layer only. The optimized output weight matrix,  $W^{{\rm out}\,\ast}$, is obtained by a minimization of a regularized 
quadratic cost function $C$. The regularization term is added to $C$ to tackle the over-fitting problem \cite{Goodfellow2016}. This cost function is given by 
\begin{align}
C \left[W^{\rm out} \right] &=\sum_{k=1}^{N_{\rm train}} \Big| W^{\rm out} {\bm r}(k\Delta t)-{\bm \Omega}(k\Delta t)\Big|^2\nonumber\\
                                            &+\gamma\, \mbox{Tr}\left(W^{\rm out} W^{{\rm out}\,T}\right)\,,
\label{RC3}
\end{align}
where ${\bm \Omega}(k\Delta t)$ is the ground truth, i.e., the PIV data. To summarize, the hyperparameters of the RCM training process are the number 
of nodes $N$, the reservoir density $D$, the spectral radius $\rho(W^r)$, the leakage rate $\alpha$, and the ridge regression parameter $\gamma>0$ of 
the regularization term of the cost function $C$. Here, we will leave $D=0.2$. {The remaining hyperparameters span a 4-dimensional space and have to 
be tuned. In the {\em training phase}, we take 100 different quadruples $\{N,\rho(W^r),\alpha,\gamma\}$ randomly from prescribed ranges and run the 
RCM for each quadruple and 10 different initial random seeds of the reservoir to optimize $W^{\rm out}$ for each of the $10^3$ training runs. The 
subsequent {\em validation phase} selects the optimal hyperparameter set as the one that gives the minimal mean squared error (MSE). This leads to the 
optimal output matrix $W^{{\rm out}\,\ast}$ for the optimal quadruple $\{N^{\ast},\rho(W^r)^{\ast},\alpha^{\ast},\gamma^{\ast}\}$.}
\begin{figure*}
\includegraphics[width=0.8\textwidth]{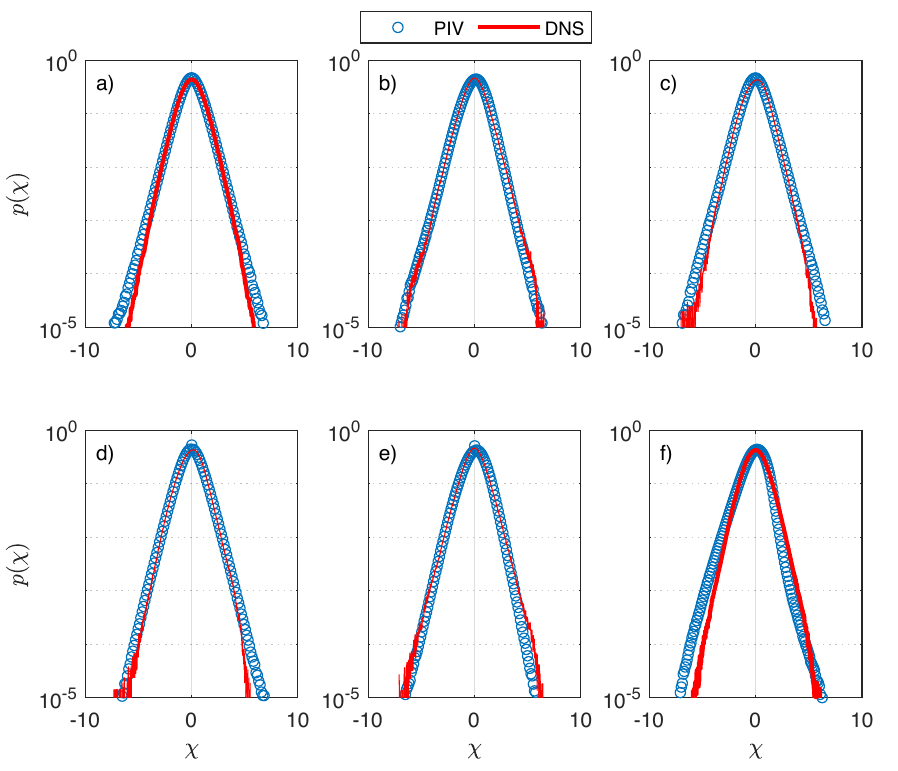}
\caption{Comparison of the probability density function of out-of-plane vorticity $\omega_z$ (a) and the five partial derivatives, which are  $\partial u_x/\partial x$ in (b), $\partial u_y/\partial x$ in (c), $\partial u_x/\partial y$ in (d), $\partial u_y/\partial y$ in (e), and $\partial u_z/\partial z$ in panel (f). The PIV experiments are at ${\rm Ra} = 5.1\times10^5$, the DNS at ${\rm Ra} = 5\times10^5$ . All quantities are normalized by their corresponding root-mean-square values. \label{fig:PDF}}
\end{figure*} 
{In the final {\em reconstruction phase}, the RCM reconstructs the vertical vorticity component field in the measurement region that can be compared with 
original unseen test data. The RCM generates synthetic time series of the intermittent vorticity component at all grid points of the measurement region 
$A$ from sparse continuously provided data at a subset of grid points of $A$. The RCM is run in a mode that is sometimes referred to as open-loop 
scenario or one-step prediction \cite{Lukosevicius2021} as the reservoir output is not fed .}

\begin{figure}
\includegraphics[width=0.45\textwidth]{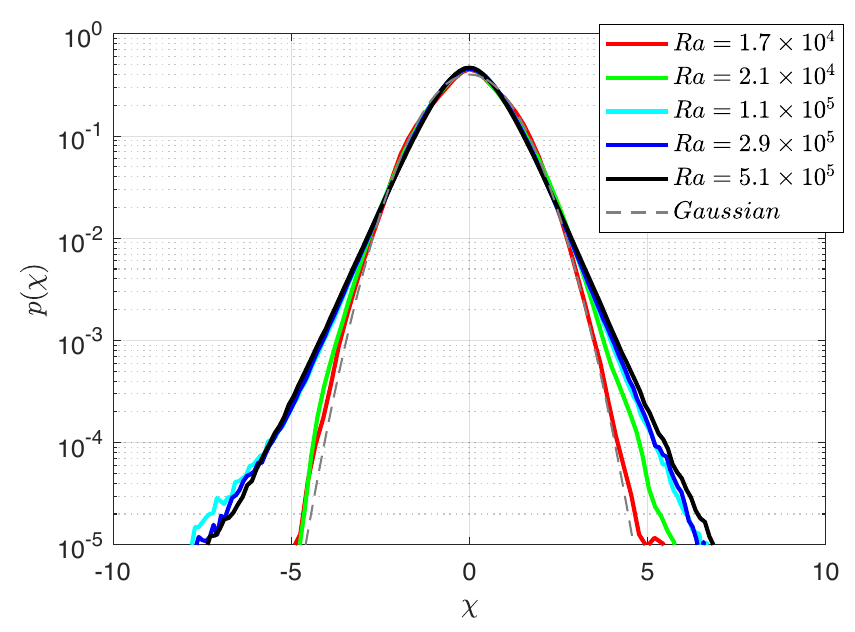}
\caption{Probability density functions (PDFs) of the out-of-plane vorticity field component obtained from different stereoscopic particle image velocimetry measurements for the Rayleigh numbers which are indicated in the legend. The gray dashed line displays the Gaussian case for reference.}
\label{allPDF}
\end{figure}
\section{Results}
\subsection{Velocity derivative statistics}
The long total measurement acquisition time of 2500 $T_{\rm ff}$ allows a good convergence of the statistics as displayed in figure \ref{convergence}. In these plots, we report results of the out-of-plane vorticity and selected individual velocity derivative components, all normalized by their corresponding root-mean-square (rms) values. We therefore define 
\begin{equation}
\chi:=\frac{\omega_z}{\sqrt{\langle \omega_z^2 \rangle_{A,t}}} \quad\mbox{or}\quad \chi:=\frac{\partial u_i/\partial x_j}{\sqrt{\langle (\partial u_i/\partial x_j)^2 \rangle_{A,t}}}\,, 
\end{equation}
with $i,j=1,2,3$. The denominators in both equations are the root-mean-square (rms) values of the corresponding quantities. The statistical convergence of the $n$-th order normalized moment follows from plots of $\chi^n p(\chi)$ versus $\chi$. The area below these curves corresponds then to the $n$-th order moment $M_n$ which is given by
\begin{equation}
M_n(\chi):=\int_{-\infty}^{\infty} \chi^n p(\chi) \,\mbox{d}\chi\,.
\end{equation}
These moments can be evaluated in a discretized approximation of this integral. The statistical convergence of moments $M_2$, $M_4$, and $M_6$ is shown in figure \ref{convergence} for the two largest experiment Rayleigh numbers $Ra = 2.9\times10^5$ and $Ra=5.1\times10^5$. A converged velocity derivative statistics implies that the tails for the largest $\chi$--values tend to decay to zero which seems to be case for the shown components. Note that $y$--axes are displayed in logarithmic units in the figure. We have also verified that the two velocity gradient tensor components which are not shown in the figure satisfy the statistical convergence criteria as well. It can be confirmed that the PIV measurements obtain sufficiently well-resolved velocity gradients in the range of accessible Rayleigh numbers. 

Figure \ref{fig:PDF} reports a direct comparison between the PDFs of 5 components of the velocity gradient tensor $A_{ij}$ and the out-of-plane vorticity component $\omega_z$ from the SPIV measurements and the DNS data from \cite{Valori2021}. One can see that the experimental results are in very good agreement with the simulation data all the way to the far tails. Again, the PDFs of the 2 missing components, $\partial u_z/\partial x$ and $\partial u_z/\partial y$, are qualitatively and quantitatively similar to the shown data.   

The PDFs of the out-of-plane vorticity of the SPIV of all series as displayed in table II are shown in figure \ref{allPDF}, respectively. One can observe that the tails of the PDFs become wider as the Rayleigh number grows, which is an indication of a transition from Gaussian to non-Gaussian intermittent velocity derivative statistics as discussed for RBC in refs. \cite{Schumacher2014,Schumacher2018,Valori2021}. We can thus conclude that this transition is also detectable in the bulk of controlled laboratory experiments at moderate Rayleigh numbers. This allows to run long-term measurements of velocity derivative statistics which is challenging in simulations where the numerical effort grows with $\Gamma^2$.

\begin{figure*}
\includegraphics[width=0.7\textwidth]{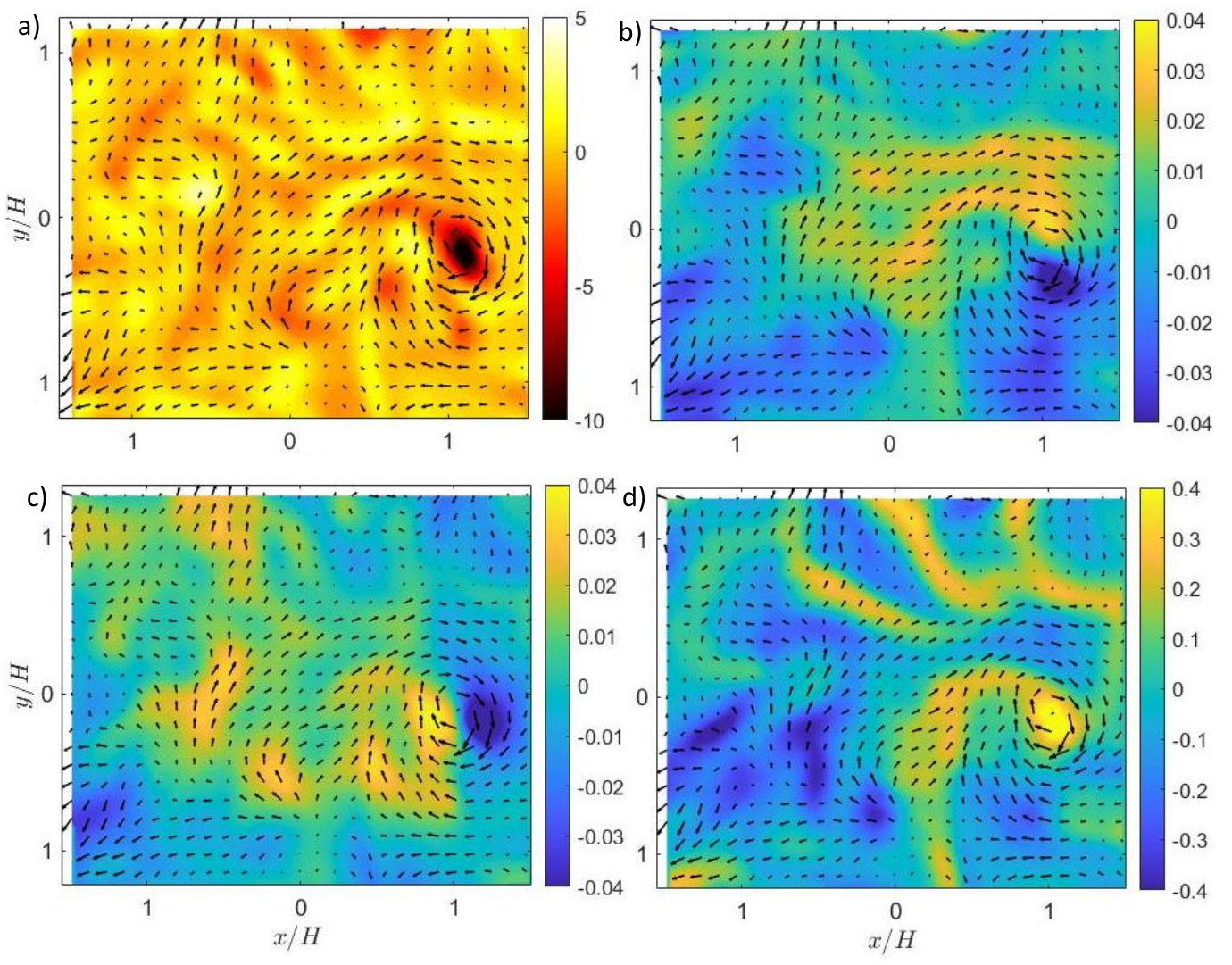}
\caption{Visualisation of an extreme event of $\omega_z$ from the SPIV measurements at $Ra = 5.1 \times 10^5$. Filled contours of $\omega_z$ in (a), $u_x$ in (b), $u_y$ in (c), and $u_z$ in (d) are shown. Color bars of the vorticity and velocity components are given in units of $t_{\rm ff}$ and $U_{\rm ff}$, respectively. Arrows show the same in-plane velocity vector field in all four panels.\label{fig_Extr}}
\end{figure*}
\begin{figure}
\includegraphics[width=0.45\textwidth]{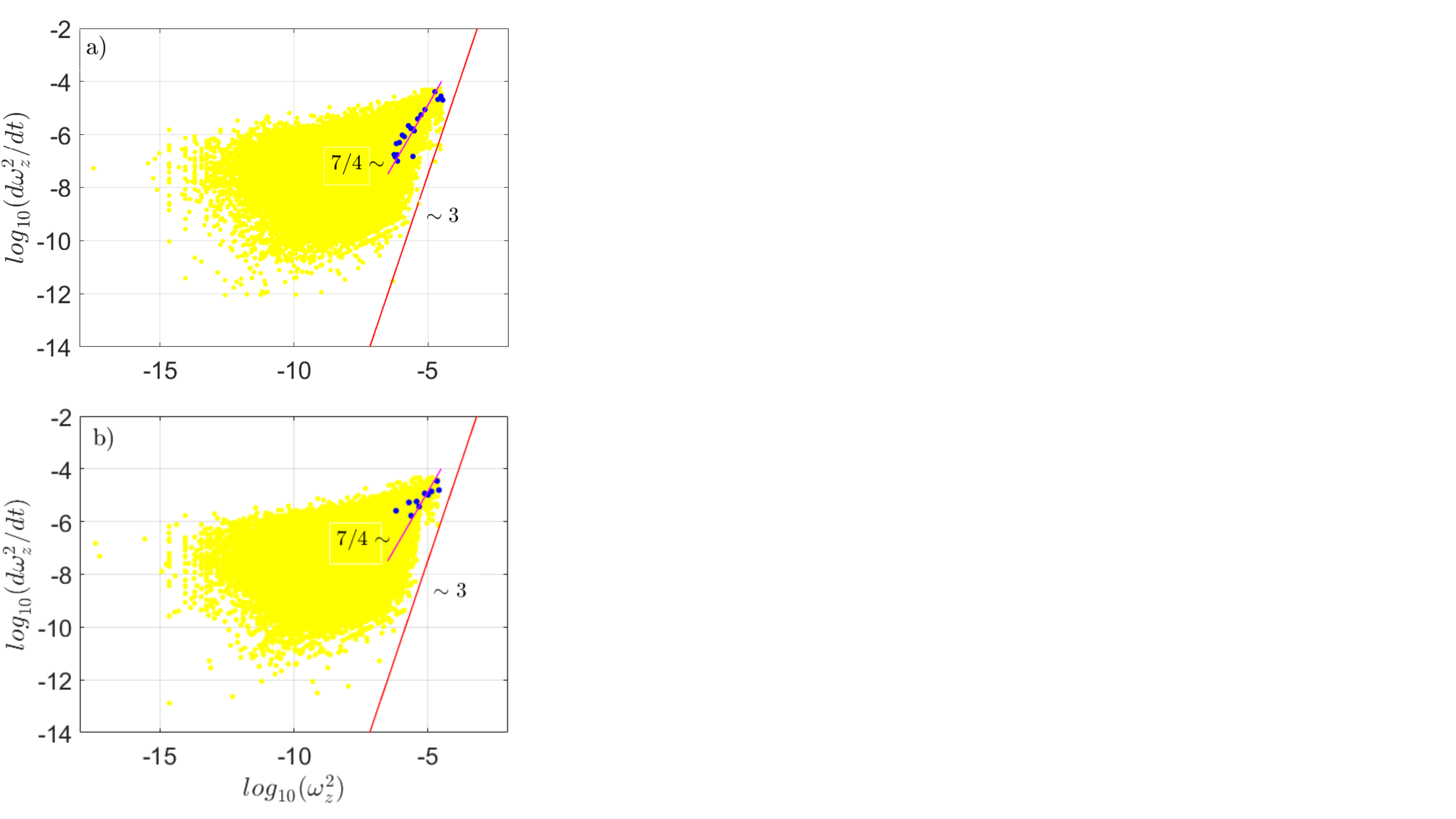}
\caption{Scatter plots of the time series, taken at the 25 grid points around extreme event position ${\bm x}_{\ast}$ for $Ra = 5.1 \times 10^5$. Panels (a) and (b) stand for the two events at $t=853.25 t_{\rm ff}$ and $t=599 t_{\rm ff}$, respectively, for which $|\omega_z|\ge 10 \omega_{z,rms}$. The squared out-of-plane vorticity growth is plotted versus squared out-of-plane vorticity. The final data points with $t<t_{\ast}$ are replotted in blue. Also indicated are the scaling exponents that follow from \cite{Lu2008}.\label{growth}}
\end{figure}

\subsection{Extreme event of out-of-plane vorticity}
An example of an extreme event of $\omega_z$ from PIV measurements at Rayleigh number $Ra = 5.1 \times 10^5$ is shown in figure \ref{fig_Extr}. We 
count an extreme event whenever the vorticity magnitude exceeds $10\omega_{z,{\rm rms}}$. In the time interval at the highest Rayleigh number, we were 
able to record two of these events. It can be expected that their frequency and excess magnitude increase when the Rayleigh number grows. This point is 
left as future work. Panel (a) of the figure shows a prominent vortex core on the right hand side of $A$. This vortex is the result of a horizontal shear in 
combination with an upward motion through the observation plane. From the corresponding velocity field, whose components are represented in panels 
(b--d) of this figure, one can suspect that a plume collision is responsible for the generation of the extreme vorticity event, as it was observed so far only in 
DNS data records \cite{Valori2021}.
Lu and Doering \cite{Lu2008} showed that the temporal growth of the enstrophy, which is given by 
\begin{equation}
\bar E(t) = \int_V \omega_i^2 dV \quad\mbox{with}\quad \omega_i({\bm x},t)=\varepsilon_{ijk} \frac{\partial u_k({\bm x},t)}{\partial x_j}\,,
\end{equation} 
in homogeneous isotropic turbulence in a triply periodic box of volume $V$, is rigorously bounded by 
\begin{equation}
\frac{d\bar E(t)}{dt} \le  \frac{27 c^3}{16\nu} \bar E(t)^3\,,
\label{enstrophy}
\end{equation} 
with $c=\sqrt{2/\pi}$. It was found that axially symmetric, colliding vortex rings maximize the enstrophy growth and that interacting Burgers vortices cause a growth $d\bar E/dt \sim \bar E(t)^{7/4}$. Even though we cannot expect that the same bound holds for a turbulent convection flow, we can probe the growth of the out-of-plane squared vorticity in the measurement region $A$.
In figure \ref{growth}, we therefore plot the temporal growth of the vorticity component, $d\omega^2_z/dt$, versus the squared vorticity, $\omega_z^2$, as 
a scatter plot for all grid points which are centered around ${\bm x}_{\ast}\in A$, the point where $\omega_z^2$ yields an extreme event at time 
$t=t_{\ast}$. In both panels the time series (which extend over $10^4$ data points) are plotted for $5\times 5$ grid points. The data right before 
$t=t_{\ast}$ are replotted in blue in both panels. In addition, the growth laws with the exponents 3 and 7/4 are also indicated by solid lines. It is seen that 
for the highest-amplitude extreme event the growth to the maximum is close to the 7/4--scaling which suggests that a vortex stretching process generates 
this event. It is also seen that the power law with a slope of 3 envelopes the data for the highest amplitudes of $\omega^2_z$ from below. Further details 
cannot be provided on the basis of the measurements due to the missing velocity derivatives and the analysis is conducted locally in contrast to 
\eqref{enstrophy}.   
\begin{figure*}
\includegraphics[width=0.73\textwidth]{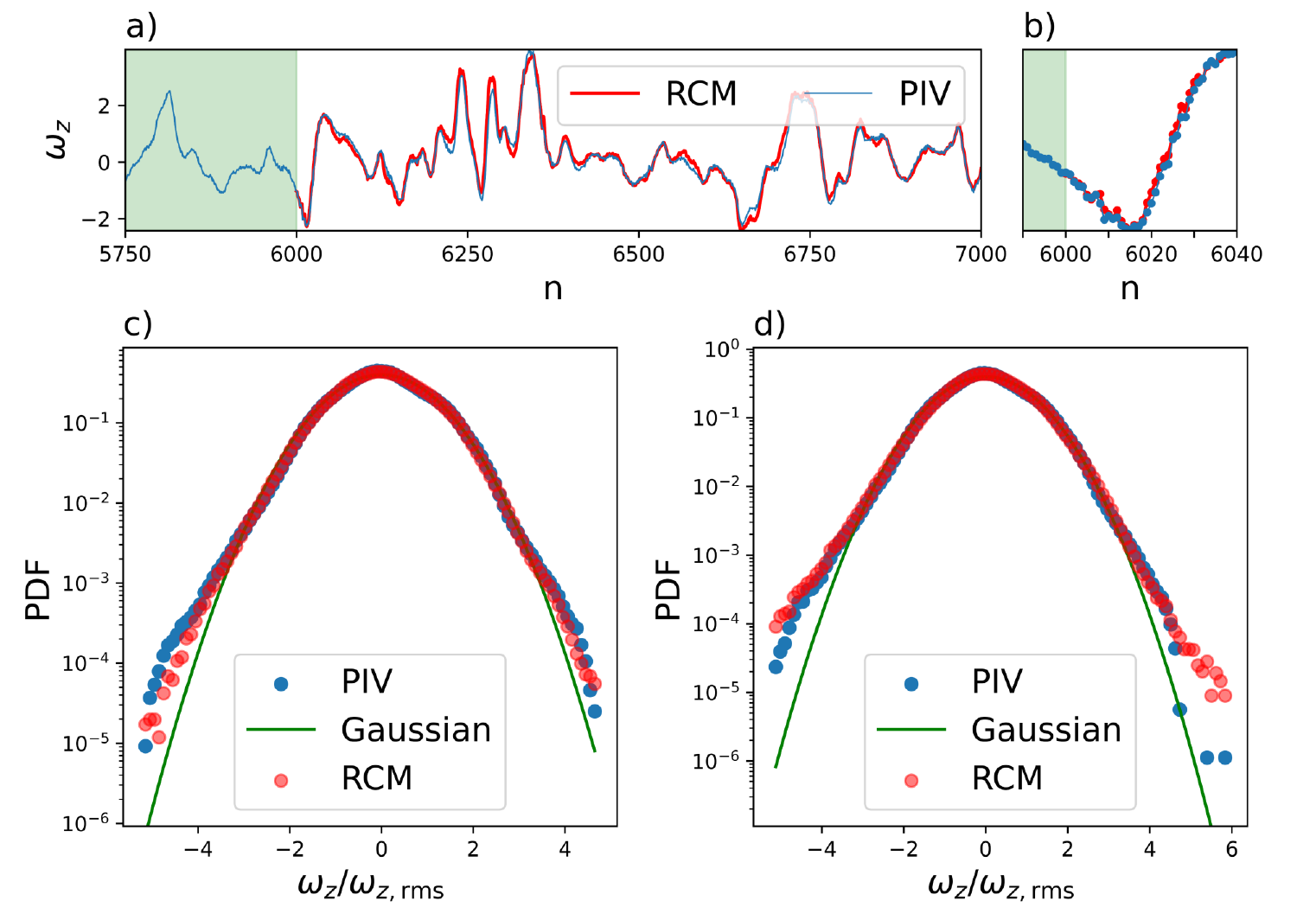}
\caption{Reconstruction of temporal evolution and statistics of the out-of-plane vorticity component $\omega_z$ by reservoir 
computing models. The Rayleigh number is ${\rm Ra}=1.7\times 10^4$. (a) Time series reconstruction example at one grid point in $A$. Integer $n$ 
corresponds to time $n\Delta t$ with $\Delta t= 0.25 t_{\rm ff}$. (b) Zoom into the first  steps of the reconstruction phase. Panel (c) shows the probability 
density function obtained from a reconstruction of correspondingly 8 grid points around each continually available data point (scenario $3\times 3$) and 
panel (d) from 48 grid points (scenario $7\times 7$).\label{fig:rcm1}}
\end{figure*} 
\begin{figure*}
\includegraphics[width=0.73\textwidth]{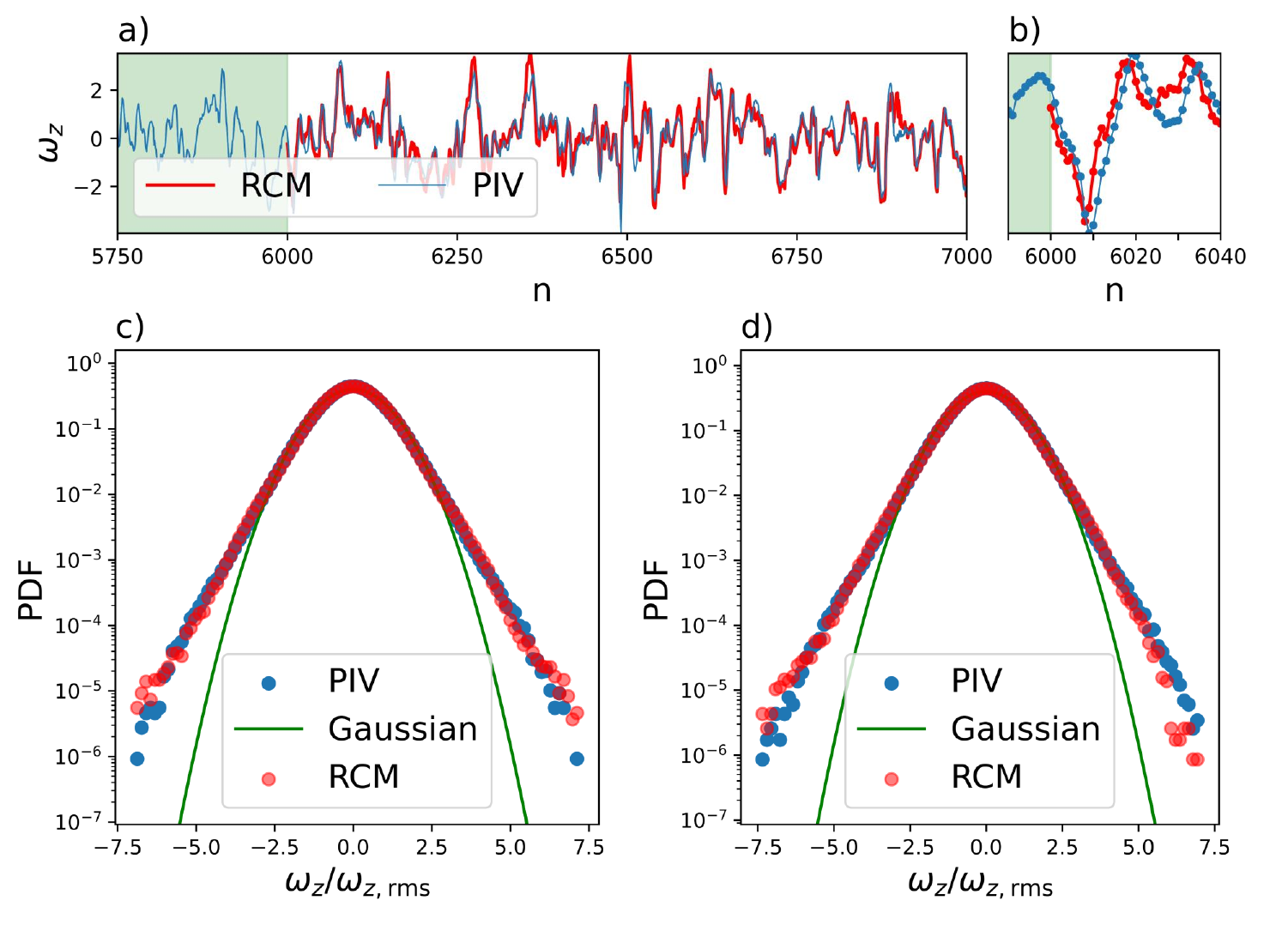}
\caption{Same as in figure \ref{fig:rcm1} for ${\rm Ra}=2.9\times 10^5$ with $\Delta t=0.25 t_{\rm ff}$.\label{fig:rcm2}}
\end{figure*} 

\subsection{Reservoir computing model for dynamical evolution of vorticity}
Finally, we apply a reservoir computing model (RCM) to reconstruct the spatial structure of $\omega_z$ in the measurement region $A$. The corresponding procedure is visualized in figure \ref{fig:sketch}(b) and follows the one  of Lu et al. \cite{Lu2017} taken to reconstruct the dynamics of the one-dimensional Kuramoto-Sivashinsky equation. We therefore provide time series of the measurements sampled on a very coarse uniform grid that covers the measurement region $A$ as seen in the sketch. The RCM is then trained to generate time series of the vorticity component at the remaining and surrounding grid points of the measurement region $A$. The procedure allows us to reconstruct the space-time dynamics of the vorticity component in $A$.

{In table \ref{tab:hyper}, we summarize the optimal hyperparameters of the RCM that are chosen after the training procedure for the time series 
prediction. All time series consist of $10^4$ PIV snapshots. The first $N_{\rm T}=5000$ snapshots were used to train the RCM model, $N_{\rm V}=1000$ 
for validation, and the remaining $N_{\rm P}=4000$ data snapshots are unseen test data in the reconstruction phase. Furthermore, we list the mean 
squared errors (MSE) in the table which follow for the training (T), validation (V), and reconstruction (R) phases. These errors are given by 
\begin{equation}
{\rm MSE}_{\xi}= \frac{1}{N_{\xi}} \sum_{n=1}^{N_{\xi}} \frac{1}{N_{\rm PIV}} \sum_{k=1}^{N_{\rm PIV}}  (\Omega_{k}(n)-\hat{\Omega}^{\rm out}_k(n))^2 \,,
\end{equation}
where ${\bm \Omega}(n)$ is again the ground truth and $\hat{\bm \Omega}^{\rm out}(n)$ the RCM output. Here, $\xi=\{{\rm T, V, R}\}$. In the reconstruction phase, the seed and hyperparameter set is used that gave the lowest MSE together with a sufficiently low Kullback-Leibler divergence \cite{Goodfellow2016} which is a second measure that was taken and defined as 
\begin{equation}
    {\rm KL}(p,q)=\sum_{\omega_z} p(\omega_z) \log\frac{p(\omega_z)}{q({\omega}_z)}\,,
\end{equation}
where $p$ and $q$ are the PDFs obtained from ground truth and RCM output in the validation phase, respectively. Note that ${\rm KL}=0$ if $p=q$ for the PDFs. We run such a hyper-parameter tuning for all 6 convection data records. The best RCM for each Rayleigh number is then used for the reconstruction task.} 
 \begin{table}
 \begin{ruledtabular}
 \begin{tabular}{lcccccc}
        &  \multicolumn{3}{c}{${\rm Ra}=1.7\times 10^4$} & \multicolumn{3}{c}{${\rm Ra}=2.9\times 10^5$} \\            
        &  $3\times 3$ & $5\times 5$ & $7\times 7$ & $3\times 3$ & $5\times 5$ & $7\times 7$\\
 \hline
 $N$              & {2523}  & {2721}  & {2687} & {2741}  & {2299}  & {2885}\\
 $\alpha$       & {0.74} & {0.14} & {0.39}      & {0.33} & {0.22}   & {0.57}\\
 $\rho(W^r)$  & {0.94} & {0.98} & {0.98}      & {0.95}  & {0.92}  & {0.93}\\
 $\gamma$    & {0.58} & {0.54} &  {0.88}   & {0.18}  & {0.02}  & {0.64}\\
 $D$               & 0.2 & 0.2 & 0.2 &  0.2 &  0.2 & 0.2\\
 ${\rm MSE}_{\rm T}$        & {0.029} & {0.024} & {0.047} & {0.098} & {0.208} & {0.29}\\  
  ${\rm MSE}_{\rm V}$      & {0.039} & {0.18} & {0.27} & {0.27} & {0.49} & {0.79}\\ 
  ${\rm MSE}_{\rm R}$ & {0.041} & {0.181} & {0.281} & {0.24} & {0.44} & {0.72} \\
  ${\rm KL}$ & {0.005} & {0.006} & {0.014} & {0.001} & {0.001} & {0.003} \\
\end{tabular}
\end{ruledtabular}
\caption{Summary of the optimal hyperparameters. Here, each run was trained individually. {Furthermore, the mean squared error (MSE) of the training 
(T), validation (V), and reconstruction (R) phases is given.} The optimal parameters were chosen for $N\in [1000,3000]$, $\rho(W_r)\in [0.9,1]$, 
$\gamma\in[0.01, 1]$, and $\alpha\in [0.1,1]$. For each of 10 different random configurations of the reservoir matrix $W^r$ 100 different random 
combinations of the hyperparameters within the given intervals were taken. {The last row displays the Kullback-Leibler divergence in the 
reconstruction phase.}\label{tab:hyper}}
\end{table}
\begin{figure*}
\includegraphics[width=0.85\textwidth]{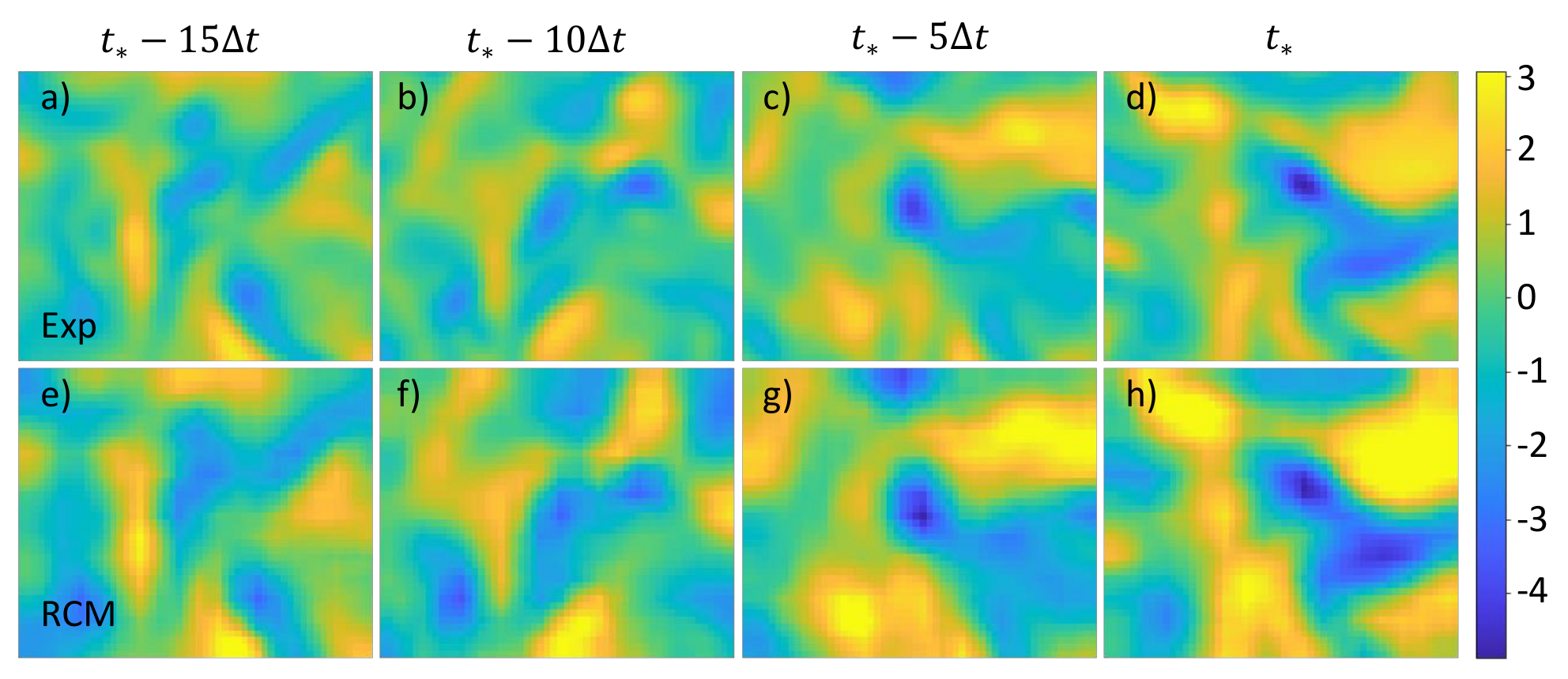}
\caption{Dynamical sequence of an extreme vorticity event (a--d) and its reconstruction by the reservoir computing model algorithm with the $3\times 3$ 
scenario (e--h). The extreme event of $\omega_z$ is at time $t=t_{\ast}$. The time interval between two snapshots is $\Delta t=0.25 t_{\rm ff}$. The 
Rayleigh number is ${\rm Ra}=2.9\times 10^5$.}
\label{fig:extreme}
\end{figure*}

Three different scenarios, namely learning the vorticity time series on $3\times 3$, $5\times 5$, and $7\times 7$ grid points around continually provided 
data points have been investigated. These three scenarios correspond to providing $11\%$, $3.5\%$ and $1.9\%$ of the total number of grid points, 
respectively. \color{black} We have conducted this analysis for the smallest and one of the largest Rayleigh numbers of our data record, ${\rm 
Ra}=1.7\times 10^4$ and $2.9\times 10^5$. The results are summarized correspondingly in figures \ref{fig:rcm1} and \ref{fig:rcm2}. Panels (a) and (b) in 
each of both figures display the reconstruction of an example of a time series taken at a specific position in $A$. We see that in both cases the 
time dependence is approximated fairly well by the reservoir computing model. Both figures are obtained for a $5\times 5$ reconstruction scenario. Panel 
(b) magnifies the initial time steps $n$ of the reconstruction phase. Panels (c) and (d) of both figures display the PDFs that result for the normalized 
vorticity component from the RCM in comparison to the experimental test data. While the velocity derivative statistics for the smallest Rayleigh number is 
still very close to a Gaussian distribution, it has crossed over to the non-Gaussian intermittent regime for the higher Rayleigh number. This becomes 
visible by the extended  tails that are also reproduced well by our machine learning algorithm. The PDFs in both panels are shown for two scenarios, 
$3\times 3$ and $7\times 7$, in each case. Note that the latter scenario implies that less information about the input is provided 
for the machine learning algorithm during the prediction phase. Only each 48$^{\rm th}$ grid point contains a partial observation. We can see again that 
the statistics in both cases are in fair agreement with the experimental results. The RCM with continually available sparse data is thus able to reconstruct  
the statistical properties of a highly intermittent out-of-plane vorticity component.

\begin{figure}
\includegraphics[width=0.48\textwidth]{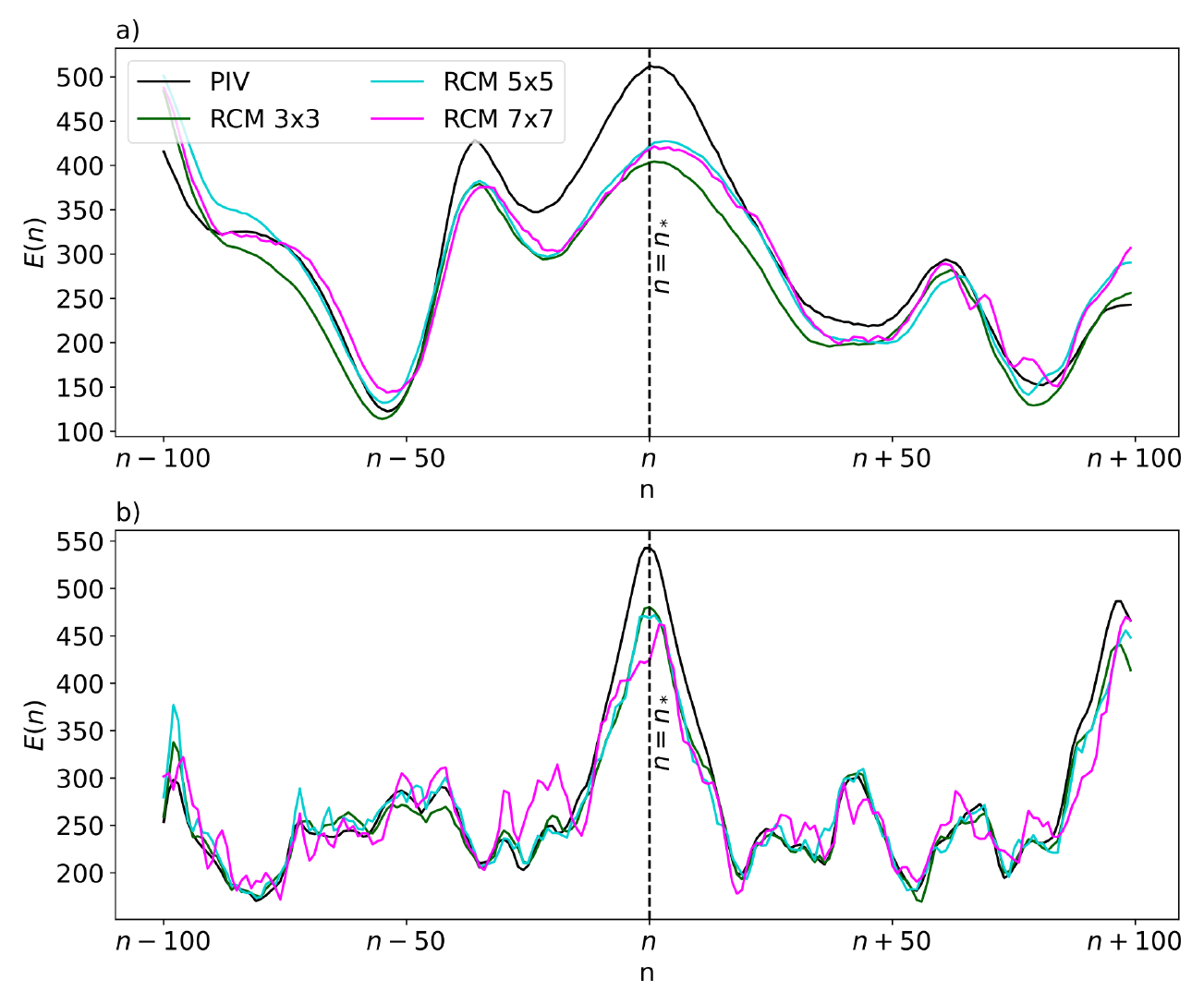}
\caption{Time series of the out-of-plane enstrophy $E(n)$ for two Rayleigh numbers, (a) ${\rm Ra}=1.7\times 10^4$ and (b) ${\rm Ra}=2.9\times 10^5$ 
which is calculated by eq. \eqref{enst}. Both plots compare the measurement data with the reservoir computing reconstructions which have been obtained 
either by the $3\times 3$, the $5\times 5$, or the $7\times 7$ scenario as indicated in the legend of panel (a). Note that argument $n$ stands for time 
$t=n\Delta t$.}
\label{fig:enstrophy}
\end{figure}
\begin{figure*}
\includegraphics[width=0.8\textwidth]{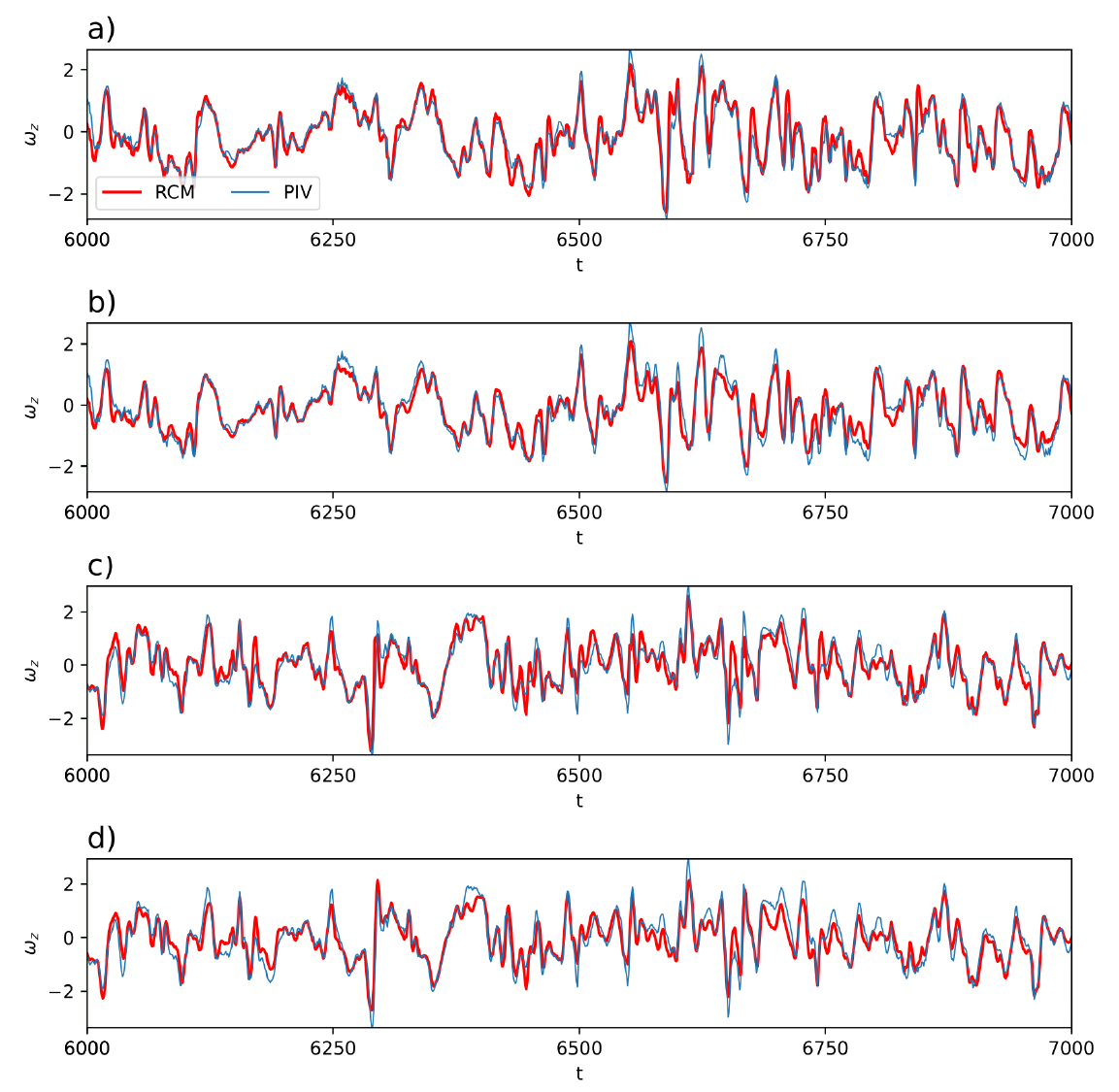}
\caption{{Generalization ability of the reservoir computing model. We show examples of reconstructed time series at one grid point in $A$. Integer $n$ 
corresponds to time $n\Delta t$ with $\Delta t= 0.25 t_{\rm ff}$. Two RCMs were trained at ${\rm Ra}=1.7\times 10^4$ and used for the reconstruction task 
for unseen data at ${\rm Ra}=1.1\times 10^5$ (a,b) and ${\rm Ra}=2.9\times 10^5$ (c,d). The first panel for each Rayleigh number shows the $3x3$ 
scenario of reconstruction (a,c) and the second panel shows the $7x7$ scenario (b,d).}}
\label{fig:rcm3}
\end{figure*}

In figure \ref{fig:extreme}, we demonstrate the capability of the trained recurrent neural network to reconstruct an extreme vorticity 
event. A typical event, which is detected at a time $t=t_{\ast}$ (that translates into snapshot number $n=n_{\ast}$) is shown here. We therefore compare a 
sequence of PIV snapshots with the corresponding model predictions. The RCM results have been composed of $3\times 3$ reconstructions. The panels 
have been subsequently smoothed by a $6\times 6$ procedure averaging. It can be seen that the figures at the corresponding times agree well with each 
other. We can thus conclude from this visual inspection that extreme or high-amplitude vorticity events can be reconstructed by the specific recurrent 
machine learning algorithm. 

In figure \ref{fig:enstrophy}, the squared out-of-plane vorticity or out-of-plane enstrophy $E(t)$ is integrated over the observation plane $A$. The quantity is given by
\begin{equation}
E(t):=\int_A \frac{\omega_z^2(t)}{2} \mbox{d}A\,,
\label{enst}
\end{equation}
and displayed for two Rayleigh numbers. The snapshot number of the global vorticity component maximum is denoted as $n_{\ast}$. The quantitative 
comparison of the 3 predictions with the experimental test data demonstrates that all 3 scenarios follow the ground truth data fairly well. {As expected, the 
deviations increase with increasing spatial sparsity of the available partial observations. This is very well visible for the larger of the two Rayleigh numbers. 
For the more quantitative analysis, we refer also to ${\rm MSE}_{\rm R}$ in Table \ref{tab:hyper}.} 

{The ability of generalization of the RCM is shown in figure \ref{fig:rcm3}. Two RCMs, one for the $3\times3$ scenario and one  for the $7\times7$ 
scenario, are trained and validated on the data with the lowest available Rayleigh number ${\rm Ra}=1.7\times 10^4$ and then tested on unseen data at 
${\rm Ra}=1.1\times 10^5$ and ${\rm Ra}=2.9\times 10^5$. The figure shows a representative comparison between the reconstructed time series on a 
selected PIV grid point and its according ground truth. The reconstructed time series follows the ground truth very well for all scenarios, which shows that 
the RCM can generalize over a certain range of experimental conditions including higher levels of turbulence.}

{We finally note here that for a reconstruction without partial observations, i.e., for a fully autonomous RCM prediction (also known as the closed-loop 
scenario), the time series at the grid points of the measurement section $A$ start to deviate after a few time steps $n$ for the highest Rayleigh numbers of 
our data record. This deviation is a known problem of RCMs \cite{Vlachas2020a}. It can be mitigated by using other recurrent neural network architectures 
such as long short-term memory networks or gated recurrent units trained with backpropagation \cite{Vlachas2020a,Vlachas2020b,Pandey2022}.}

\section{Summary and outlook}
Our present work was motivated by two major scientific objectives: (1) the detailed analysis of the intermittent statistics of velocity derivatives in the bulk of 
a turbulent Rayleigh-B\'{e}nard convection flow in air by means of stereoscopic particle image velocimetry measurement including the monitoring of 
high-amplitude events of the out-of-plane vorticity component and (2) the machine--learning--assisted reconstruction of the dynamical evolution and 
statistics of the small-scale velocity derivatives including the reconstruction of extreme or high-amplitude events. Therefore, the moderate Rayleigh 
numbers were varied over an order of magnitude, in a range for which the statistics of the spatial velocity derivatives (and thus of the vorticity 
components) goes over from Gaussian to non-Gaussian, as discussed in our recent direct numerical simulations \cite{Valori2021}. {This transition in the 
derivative statistics was detected by both, the experiments and the subsequent machine learning algorithm which is based on a recurrent neural network 
in the form of a reservoir computing model. The latter was trained on the experimental PIV data.}

As in most laboratory experiments and in contrast to direct numerical simulations the three velocity components of the turbulent convection flow were detectable in a horizontal section only and not fully resolved in the whole volume. Exceptions are high-resolution experiments which are practically almost as expensive in their postprocessing as fully resolved direct numerical simulations \cite{Westerweel2013,Schanz2016}. SPIV allows us to reconstruct 7 out of the 9 components of the velocity gradient tensor $M_{ij}$ together with the out-of-plane vorticity component $\omega_z$. In many situations, such as field measurements,  the time series data are taken at sparsely distributed locations. These practical constraints suggest the application \color{red}of \color{black} (recurrent) machine learning algorithms. They can process sequential data, make predictions on the dynamics, and thus add missing dynamical information on the turbulence fields. In the second part of this work, we had exactly such a proof-of-concept in mind when investigating the reconstruction capabilities of the applied reservoir computing model particularly those of extreme or high-amplitude vorticity events.   

The studies can be extended in several directions. One direction would be a combination with temperature measurements close to the boundary layer, 
such that strong plume detachments can serve as precursors for extreme dissipation or vorticity events in the bulk of the convection layer. This idea has 
been developed in ref. \cite{Valori2021} on the basis of fully resolved DNS. {In this case, super-resolution generative adversarial networks can reconstruct 
the temperature field from coarse measurements close to the wall. A similar path has been taken for near-wall velocity measurements in ref. 
\cite{Guemes2021}.} A further direction consists of the design of generative algorithms that produce time series of the missing 2 velocity derivative 
components complemented by the existing statistical symmetries in the flow at hand \cite{Sondak2019,Karniadakis2021}. In this way, 
machine--learning--assisted measurements of the kinetic energy and thermal dissipation rates would be possible without the usage of tomographic 
techniques. Studies in this direction are currently underway and will be reported elsewhere.     
 
\acknowledgements
The authors would like to thank Alexander Thieme for the support with the experiments and Florian Heyder for helpful remarks. The work of VV was partly supported by Priority Programme DFG-SPP 1881 on Turbulent Superstructures of the Deutsche Forschungsgemeinschaft (DFG). Her work is currently supported by the Marie Curie Fellowship of the European Union with project number 101024531. The work of RK is supported by project SCHU 1410/30-1 of the DFG. Training and prediction of the machine learning algorithm were carried out on up to 8 CPUs of the compute cluster Makalu at Technische Universit\"at Ilmenau.

\bibliographystyle{unsrt}
\bibliography{Biblio}

\end{document}